\documentclass[a4paper]{article}

\usepackage{INTERSPEECH2020}
\usepackage{multicol}
\usepackage{graphicx}
\usepackage{multirow}

\PassOptionsToPackage{hyphens}{url}\usepackage{hyperref}

\allowdisplaybreaks

\title{GAN Vocoder: Multi-Resolution Discriminator Is All You Need}
\name{Jaeseong You, Dalhyun Kim, Gyuhyeon Nam, Geumbyeol Hwang, Gyeongsu Chae}
\address{
  MoneyBrain Inc., Seoul, Korea}
\email{\{jaeseongyou, torch, ngh3053, goldstar, gc\}@moneybrain.ai}

\begin{document}

\maketitle
\begin{abstract}
Several of the latest GAN-based vocoders show remarkable achievements, outperforming autoregressive and flow-based competitors in both qualitative and quantitative measures while synthesizing orders of magnitude faster. In this work, we hypothesize that the common factor underlying their success is the multi-resolution discriminating framework, not the minute details in architecture, loss function, or training strategy. We experimentally test the hypothesis by evaluating six different generators paired with one shared multi-resolution discriminating framework. For all evaluative measures with respect to text-to-speech syntheses and for all perceptual metrics, their performances are not distinguishable from one another, which supports our hypothesis.
\end{abstract}

\section{Introduction}

Neural vocoder (i.e. neural network models that transform from speech features to raw waveform) has been one of the most active fields of study in the domain of speech synthesis. Modeling raw audio is essential in vocoding, and it poses many difficulties: (1) a digital speech signal tends to be a sequence of extended length, which brings in complications of computation and memory cost, (2) a speech signal is extremely shift-variant, rendering a straightforward reconstruction loss difficult, (3) the receptive field of a model should be extensive to catch long dependencies in audio that are essential in generating low frequency signals, (4) reconstructing high frequency components, albeit being a key to fidelity of synthesized speech, is not prone to modeling due to their noise-like nature, and (5) the temporal upsampling from mel spectrogram to waveform results in patternized artifacts, which are particularly audible.

Regardless of the hurdles, many effective vocoder models have been proposed. The initial landmark success was achieved by the autoregressive models \cite{WaveNet, WaveRNN}. Their causal inductive bias is elegantly in sync with the nature of speech signals, and they minimize the data log likelihood directly. With the powerful model assumption and the solid theoretical ground, autoregressive models had been the best-performing vocoders for a long time. Their subsequent knowledge-distilled variants are mostly geared to lessen the computational overhead for training and inference \cite{ClariNet, PWNet}.

To avoid the sample-by-sample causal inference or the use of teacher models, parallelizable statistical models have been devised. Particularly notable are the flow-based vocoders \cite{WaveGlow, FloWaveNet} that shape a simple prior distribution into a complex data distribution conditioned on a mel spectrogram using a chain of invertible operations. Although the constraint of invertibility allows direct modeling of the data log likelihood, it also forces the use of element-wise affine transformation, which in compensation for its simplicity, necessitates a massive WaveNet-like module repeating many times. More recently, diffusion-based models \cite{WaveGrad, DiffWave} achieve impressive negative log-likelihood and mean opinion score (MOS) by letting the network learn how to denoise from a Gaussian noise to a target signal in an interactive synthetic process.

\begin{figure*}
  \centering
  \includegraphics[scale=0.24]{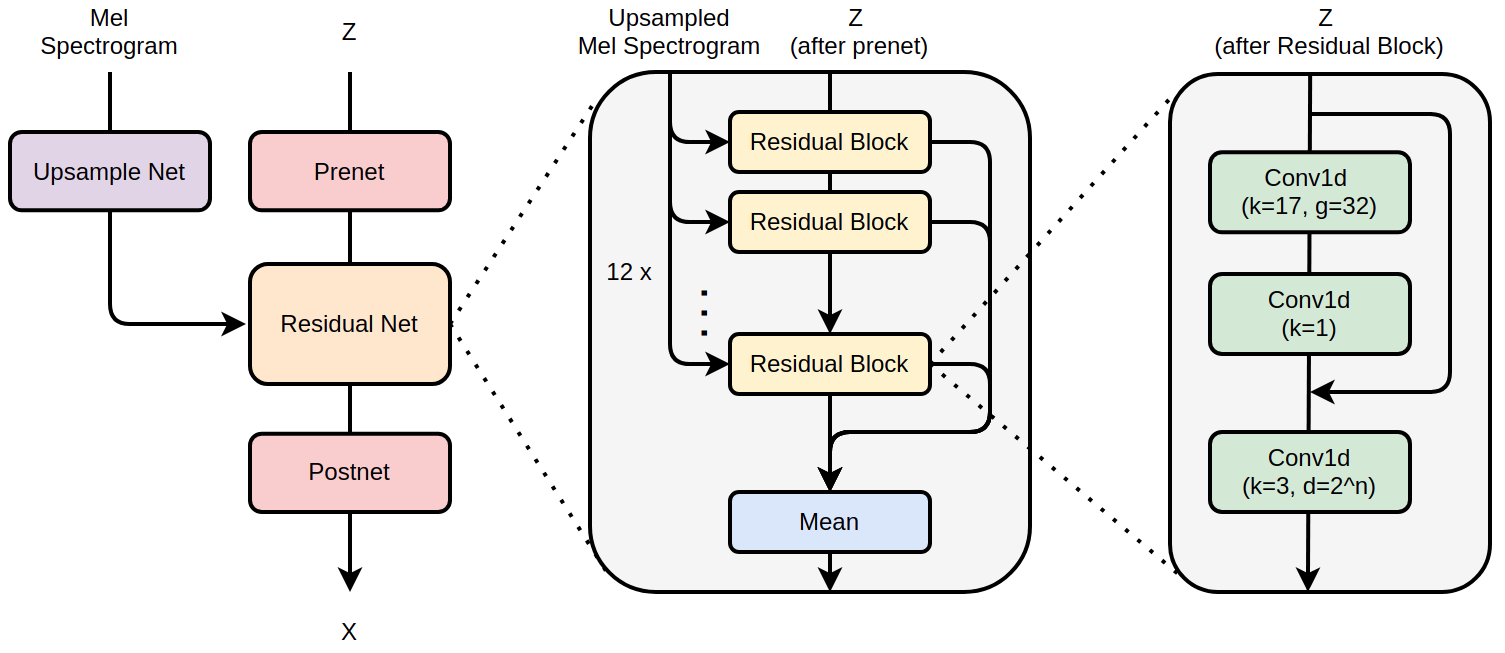}
  \setlength{\belowcaptionskip}{-15pt}
  \caption{Proposed architecture: overall generator (left), residual net (middle), and residual block (right). With the first convolution of a large kernel size and the third convolution of exponentially increasing dilation, the receptive field grows to 8369 samples by the 12th block. The residual net wires WaveNet-like skip connections from every residual block to the 1$\times$1 convolution postnet.}
  \label{fig:arc}
\end{figure*}

On the other hand, Generative Adversarial Networks (GANs) \cite{Goodbro}, although its strong potential had been demonstrated in the field of image synthesis, did not show a viable performance as a vocoder until recently. Early attempts \cite{WaveGAN, cWaveGAN} suffered from severe upsampling artifacts. MelGAN \cite{MelGAN}, the first effective GAN vocoder, too could not reach the MOS values of the flow-based models and the autoregressive models.

However, recent GAN vocoders made breakthrough improvements \cite{PWGAN, UMGAN, PWGANPW, VocGAN, HiFi-GAN}, often exceeding the previous state-of-the-art WaveNet on various metrics both qualitatively (e.g. MOS) and quantitatively (e.g. mel cepstral distortion) while being orders of magnitude faster to train and to infer. Their performances are so remarkable that, given a single-speaker dataset of sufficient quantity and consistent quality, one might even claim that the problem of vocoding has been solved.

How do the latest GAN vocoders perform so eminently? For each of the models, researchers underline many different key factors for its success: generator architecture, discriminator architecture, loss function composition, and various training tricks. In this paper, we question their validity as decisive elements. We instead hypothesize that the most important underlying common factor behind the recent successes of GAN vocoders is the multi-resolution discriminating framework, not necessarily the aforementioned details. To demonstrate this experimentally, we pair each of different generators of recent GAN vocoders as well as our proposed generator with the discriminator of HiFi-GAN \cite{HiFi-GAN} and report the outcomes in detail. We encourage readers to listen to audio samples on the companion webpage\footnote{\url{https://moneybrain-research.github.io/gan-vocoder}.}, where all the synthesized audios we use for the MOS test are available.

\section{Related Works}

We take the multi-resolution discriminating framework of HiFi-GAN as the baseline for three reasons: (1) structural simplicity: since its multi-resolution discriminating framework requires none but the final waveform prediction from the generator (no additional intermediate generator features are needed unlike in \cite{VocGAN}), it is compatible with many different forms of generator, (2) performance: given the reported metrics in \cite{HiFi-GAN}, it is currently one of the best vocoder models, outperforming WaveNet and WaveGlow in both ground truth (GT) reconstruction and text-to-speech (TTS) synthesis, and (3) official repository: HiFi-GAN publicly shares a well-maintained official implementation \cite{HiFi-GAN_git}.

The HiFi-GAN generator takes a mel spectrogram input and expands it while reducing the number of channels by alternating transposed-convolution-based upsamplings and multi-path residual blocks called multi-receptive field fusion (MRF). MRF processes multiple patterns of various lengths from the input to the residual block at once in a parallelized manner. As to both generator and discriminator implementations for our experiment, we refer to the official project repository \cite{HiFi-GAN_git}.

MelGAN was the first demonstration that a GAN vocoder can achieve a performance that comes close to that of WaveGlow \cite{WaveGlow}. For every level in the upsampling process, it pairs a residual stack of three dilated convolution layers and a transposed convolution. This design now serves as a standard baseline for subsequent GAN vocoders due to its simplicity and effectiveness. For our experiment, the MelGAN generator of the official project repository is utilized \cite{MelGAN_git}.

Parallel WaveGAN (PWGAN) and its enhanced variation that applies a perceptually weighted mask to spectrogram when computing losses \cite{PWGANPW} generate audio by taking a Gaussian noise. This is a critical difference from four existing GAN vocoders in Table~\ref{tab:gens} (except for ours) where the output results from upsampling a mel spectrogram without employing a latent variable. In addition, PWGAN employs a strictly WaveNet-like but non-causal architecture as a generator, contrasting against more MelGAN-like vocoders such as VocGAN and Universal MelGAN (UMGAN). We use an unofficial but well-maintained implementation \cite{PWGAN_git}.

The overall architecture of the UMGAN generator is similar to that of MelGAN, consisting of residual stacks and transposed convolution upsampling layers. But within each of the residual blocks, UMGAN adopts a gated activation unit (GAU), similarly to WaveNet and PWGAN. There is no available implementation of UMGAN so we construct its generator ourselves as reported in \cite{UMGAN}. For missing details, we refer to MelGAN and MultiBand MelGAN \cite{MBMelGAN}, which are the baseline models of UMGAN.

VocGAN is different from the other state-of-the-art GAN vocoders in that it is originally designed to utilize both the final output and the intermediate outputs of the generator as the target of discrimination. It has a longer upsampling sequence of 4$\times$, 4$\times$, 2$\times$, 2$\times$, 2$\times$, 2$\times$ than the usual MelGAN style of 8$\times$, 8$\times$, 2$\times$, 2$\times$ and requires auxiliary conditioning from a mel spectrogram for the last four residual blocks. Since there is no available official repository, an unofficial implementation \cite{VocGAN_git} is adopted for our experiment.

\begin{table*}[th]
  \caption{Summary of the six models employed for our experiment. The reported numbers of parameters concern only the generators, and the training speeds are recorded when training each generator paired with the HiFi-GAN V2 discriminating framework. We average the elapsed time from the beginning to the stoppage at 700k steps per batch. The inference speed is computed by averaging the elapsed time of synthesizing 150 evaluation GT mel spectrograms.}
  \label{tab:gens}
  \centering
  \begin{tabular}{clllll}
    \toprule  
             & Parameter    & Training speed & Inference speed  & Input type      \\
    \midrule
    HiFi-GAN & 928,514      & .61 s / batch  & .0058 s / sample & Mel spectrogram \\
    MelGAN   & 4,266,050    & .58 s / batch  & .0033 s / sample & Mel spectrogram \\
    PWGAN    & 1,320,442    & .82 s / batch  & .0098 s / sample & Gaussian noise (+ mel)  \\
    UMGAN    & 93,077,506   & 1.09 s / batch & .0035 s / sample & Mel spectrogram \\
    VocGAN   & 4,715,698    & .61 s / batch  & .0085 s / sample & Mel spectrogram \\
    Ours     & 1,209,590    & .78 s / batch  & .0048 s / sample & Gaussian noise (+ mel) \\ 
    \bottomrule
  \end{tabular}
\end{table*}

\section{Proposed Architecture}

We propose a generator architecture of our own. Each residual block is constructed similarly to \cite{Axial}, which efficiently widens the receptive field by conjoining two convolution layers that process patterns over temporal axis and frequency axis respectively. In our proposed method, the first layer performs group-wise convolution, enabling a very large kernel size with a small number of parameters. The second convolution layer has a standard kernel size of 3 but with an exponentially increasing dilation. In between the two layers, we introduce 1$\times$1 convolution that mixes up the information between channels. Every convolution layer output is followed by the non-linearity of LeakyReLU with slope of 0.1. The residual connection is made between the block input and the output of the in-between 1$\times$1 convolution layer. The architecture is illustrated in Figure~\ref{fig:arc}.

Mel spectrogram upsampling is done by alternating nearest-neighbor upsampling and 1D convolution layer with the kernel size of 3. We use nearest-neighbor upsampling over transposed convolution as \cite{Dolby, VDVAE} report various advantages (e.g. less distortion in the frequency domain, less checkerboard artifacts, and better preservation of information from low resolution). Upsampling is done in 8$\times$, 8$\times$, 2$\times$, 2$\times$ following the MelGAN procedure, and the upsampled mel is added element-wise to the input to every block.

The purpose behind the proposed architecture is to reinforce the diversity of the six generators so as to further test the generality of the multi-resolution discriminating framework. The proposed generator can be interpreted to be a mixture of varying generators; it takes a MelGAN-like simple residual architecture, WaveNet-like skip connection from all the intermediate layers to the ultimate 1$\times$1 convolution layer, and a PWGAN-like latent-variable-dependent process with mel spectrogram as a guiding condition.

\section{Experiments}

We use the LJSpeech dataset \cite{LJ}, consisting of 13,100 English utterances of a single female speaker with a total duration of approximately 24 hours. It is split into a training set of 12,950 files and a validation set of 150 files. The latter is not used for training, preserved only for evaluation. The audio format is 16-bit PCM with a sample rate of 22.05 kHz. The dataset is employed as is without any modification. 

Aside from the varying generator, we strictly follow the V2 configuration of HiFi-GAN with respect to the data processing procedure, the discriminating mechanism, and the loss composition, as proposed in \cite{HiFi-GAN} and as specified in \cite{HiFi-GAN_git}. We adopt the V2 configuration over V1 since its reported performance is not much different from that of the optimal V1, it is a straightforwardly smaller version of V1 (V3 introducing more fundamental changes), and it trains more quickly than V1.

80-band mel spectrograms are computed with the FFT size, the window size, and the hop size set to 1024, 1024, and 256, respectively. We use the AdamW optimizer \cite{AdamW} with $\beta_{1}$ = 0.8, $\beta_{2}$ = 0.99, and the weight decay of $\lambda$ = 0.01. The learning rate decays by 0.999 every epoch, starting from 2e-4. The batch size is set to 16. Since the five existing GAN vocoders \cite{MelGAN, PWGAN, UMGAN, VocGAN, HiFi-GAN} originally report a wide range of numbers of training steps from 400k to 2.5M, we set the stoppage to an approximate midpoint (700k steps) for all the six models of our experiment.

To evaluate the audio quality, 20 sentences are randomly selected out of the 150 evaluation samples disjoint from training. We then compute the mel spectrograms of the chosen utterances and have each of the vocoders synthesize them back to raw waveforms. In addition to the GT mel spectrogram-based vocoding, we use Tacotron 2 \cite{Tacotron2} to test the performances of the six generators against the TTS output. To this end, the Pytorch example implementation of Tacotron 2 \cite{DLE} and its pretrained weights \cite{DLE_weight} are employed.

Given 6 vocoders, 20 GT-based syntheses, and 20 TTS-based syntheses, 240 synthesized utterances are obtained in total. We then add 100 of the 150 evaluation samples (these are original human utterances with no vocoding involved) to the set; the samples corresponding to the 20 sentences used for GT- and TTS-based syntheses are first included, and the remaining 80 are randomly chosen. The resulting final test set thus contains 340 utterances in a random order. All the audio clips are normalized to the equal loudness of -21 dB to prevent the influence of audio volume differences.

Each of the randomly shuffled 340 samples is assigned to 30 unique subjects via Amazon Mechanical Turk (AMT) \cite{AMT}. The participants are asked to evaluate the audio files regarding their naturalness (i.e. human-sounding) on a five-point Likert scale (1: Bad, 2: Poor, 3: Fair, 4: Good, 5: Excellent). The total number of raters who participated is 189. To avoid any response bias, we use the AMT template of audio naturalness as is with no modification and accept all the collected responses without rejection on a first-come-first-serve basis.

In addition to the qualitative method, the mel cepstral distortion (MCD) \cite{MCD} is utilized to evaluate the models by a quantitative measure. For all the pairs of the GT and the corresponding synthesized utterances, we represent them as mel-frequency cepstral coefficients (MFCC). Additionally, dynamic time warping is applied to align the MFCC sequences of TTS-based syntheses to their GT pairs (the GT-based syntheses are aligned to their correspondences by default) as in \cite{MCD_DTW}. We use 13 coefficients excluding the first coefficient. Before computing MCD, all audio clips are normalized to have mean of 0 and standard deviation of 1 to minimize the volume differences.

\section{Results}

In Table~\ref{tab:MOSMCD}, the MOS scores of the six generators are distributed in an extreme vicinity of one another for both TTS-based and GT-based syntheses. For either case, the confidence interval of the best performing model (ours) overlaps with that of the worst performing model (PWGAN in $MOS_{GT}$ and MelGAN in $MOS_{TTS}$) despite a large number of collected answers (600) per model; no single model is therefore significantly outperforming or underperforming.

\begin{table}[th]
  \caption{Comparison of generators by MOS and MCD. The results are synthesized from both GT mel spectrograms (left) and Tacotron2 TTS outputs (right). GT* denotes MOS tests on original (not synthesized) audio files.}
  \label{tab:MOSMCD}
  \centering
  \begin{tabular}{clll}
    \toprule  
             & $MOS_{GT}$                   & $MOS_{TTS}$       \\
    \midrule
    HiFi-GAN & 3.985 $\pm$ .065             & 3.657 $\pm$ .082  \\
    MelGAN   & 3.995 $\pm$ .065             & 3.580 $\pm$ .079  \\
    UMGAN    & 3.997 $\pm$ .067             & 3.615 $\pm$ .077  \\
    VocGAN   & 3.980 $\pm$ .066             & 3.700 $\pm$ .076  \\
    PWGAN    & 3.938 $\pm$ .074             & 3.650 $\pm$ .079  \\
    Ours     & \textbf{4.028} $\pm$ .065    & \textbf{3.708} $\pm$ .075 \\
    GT*      & \multicolumn{2}{c}{4.132 $\pm$ .028}             \\
    \midrule
             & $MCD_{GT}$                   & $MCD_{TTS}$       \\
    \midrule
    HiFi-GAN & 6.425 $\pm$ .050             & 12.937 $\pm$ .145 \\
    MelGAN   & 6.160 $\pm$ .050             & 12.850 $\pm$ .138 \\
    UMGAN    & \textbf{5.365} $\pm$ .059    & \textbf{12.824} $\pm$ .138 \\
    VocGAN   & 6.500 $\pm$ .050             & 13.023 $\pm$ .140 \\
    PWGAN    & 6.196 $\pm$ .055             & 12.882 $\pm$ .137 \\
    Ours     & 6.680 $\pm$ .052             & 13.013 $\pm$ .136 \\
    \bottomrule
  \end{tabular}
\end{table}

In the case of the TTS-based syntheses, the MCD values in Table~\ref{tab:MOSMCD} show a similar pattern; all the models are once again within the confidence intervals of one another. Therefore, in an actual TTS application where utterances are inferred based on unseen texts, the synthesis quality will be similar both quantitatively and qualitatively amongst the six vocoders. The MCD scores of the GT-based syntheses, however, are the only case of the four types of measures, in which the models show statistically significant differences.

\begin{figure}
  \centering
  \includegraphics[scale=0.2]{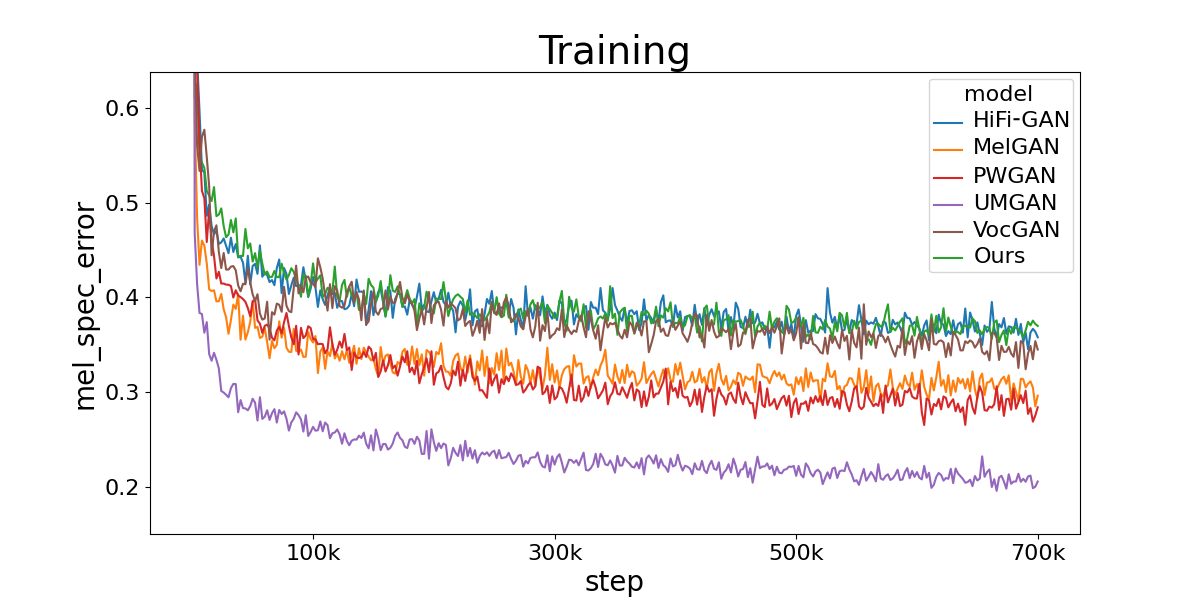}
  \includegraphics[scale=0.2]{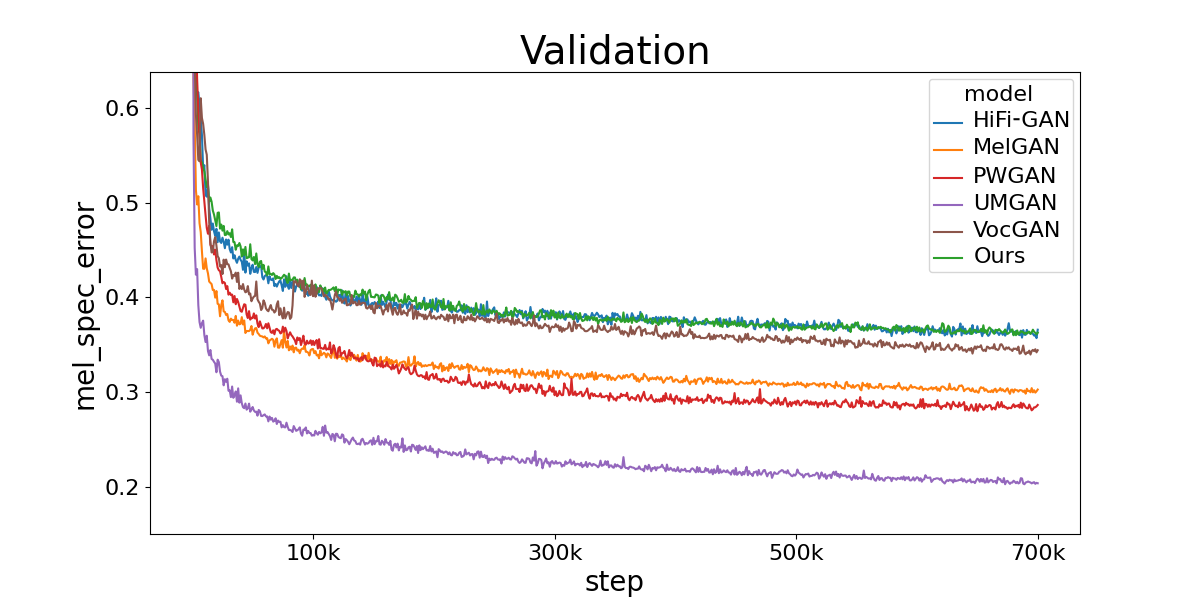}
  \setlength{\belowcaptionskip}{-15pt}
  \caption{L1 reconstruction loss on mel spectrograms over time; training (top) and validation (bottom). X-axis unit is batch step.}
  \label{fig:overtime}
\end{figure}

The differences in the quantitative performance of the GT-based synthesis amongst the six generators are better illustrated in Figure~\ref{fig:overtime} (since the mel spectrogram error is computed as an L1 loss between a GT sample and the corresponding reconstruction, one can interpret it as the model performance on GT mel spectrogram over time). Three groups are observed: (1) UMGAN does consistently best from early on for both training and validation errors, (2) MelGAN and PWGAN follow, and (3) HiFi-GAN, VocGAN, and our proposed model constitute the last group. Although the heaviest UMGAN shows the best performance, the subsequent models' performances are not in the order of the numbers of parameters. We thus observe no clear factor that explains the differences in $MCD_{GT}$.

\section{Discussion}

Architecture-wise, the six generators differ in many aspects: MRF of HiFi-GAN, axial residual block and nearest-neighbor upsampling of our proposed model, UMGAN's GAU residual block adapted to MelGAN-like architecture, and a longer upsampling sequence and auxiliary mel conditioning of VocGAN, to mention a few. Input-wise, PWGAN and our proposed model take a noise latent and shape it into a waveform while others upsample directly from a mel spectrogram. The capacities of the six generators greatly vary as well. Given these considerable differences, it is unexpected that the vocoders achieve highly similar performances, failing to demonstrate distinguishable scores for three out of the four metrics. Noticeable is the original HiFi-GAN generator neither performing prominently nor training/inferring quickly. It shows no sign of advantage from being paired with its intended discriminator pair.

Our conjecture is that the important common traits amongst the six generators are (1) effective capturing of long dependencies in the audio and (2) dense wiring that exposes every layer closely to the gradients from the final loss. As long as these primary conditions are met, other architectural details do not seem to affect the vocoded outcome significantly. While recent studies closely link generators and discriminators and explicate their interactions to minute details, the experimental results suggest a different possibility: given a well-established multi-resolution discriminating framework and an operating generator, the architectural details might not be critical factors of the vocoder performance.

Note that the original discriminator of UMGAN determines real/fake based on spectral representations of GT utterances and predictions as well as their raw waveforms. VocGAN requires intermediate and final waveform representations from its generator for discrimination. Although paired with a discriminating system that examines only the final output waveform here, both UMGAN and VocGAN generators achieve qualitative and quantitative scores just as high as other generators. The implication here is that the multi-resolution discriminating framework is applicable even across various forms of representations: raw waveform, spectral representation, and intermediate network features.

From the experiment, we can see that the light generators of HiFI-GAN, PWGAN, and our proposed one show comparable performances to those of heavier models such as MelGAN, VocGAN and even UMGAN that is 9 to 10 times larger. In addition, the MelGAN generator, despite being an earlier and simpler instance, does just as well as more modern designs. This observation implies the fact that the capacity and the wiring of the generator have already been sufficiently powerful since MelGAN. Therefore, increasing the number of channels, the number of layers, or the number of blocks might not be a key to further advancing the vocoder performance.

We believe that the findings above are applicable beyond vocoding. Given the powerful generality of the multi-resolution discriminating framework, it is highly likely to improve various GAN-based speech tasks such as speech synthesis, voice conversion, speech enhancement, and singing synthesis among others.

\section{Conclusions}

In this work, we pair the HiFi-GAN multi-resolution discriminating framework with five different generators of the latest GAN vocoders. Our proposed generator is added to the set to enhance the overall variety of the six generators’ architectural characteristics. MOS and MCD evaluations are performed on GT-based and TTS-based syntheses of the vocoders, and in three of the four metrics, we observe no sign of statistical difference in their performances despite their widely different structures. This finding questions what really matters in creating a high-fidelity speech signal from a mel spectrogram in a GAN setting and supports the hypothesis that the deciding factor lies in the multi-resolution discriminating framework. From the experiment, we make three noteworthy observations: (1) the multi-resolution discriminating framework can be generalized across different generators of varying design philosophy, (2) the multi-resolution discriminating framework can be generalized across various forms of targets (i.e. spectrogram, raw waveform, or intermediate representation) to be discriminated, and (3) the capacity and the wiring of the generator have already been sufficiently powerful since MelGAN. Observing the strong generality of the multi-resolution discriminating framework, we expect it to be effective in various other GAN-based speech tasks.

\bibliographystyle{IEEEtran}

\bibliography{mybib}


\end{document}